\documentclass[%
    reprint,superscriptaddress,
    nofootinbib,
    amsmath,amssymb,
    aps,pra,
    floatfix,
]{revtex4-2}
\pdfoutput=1 % Make sure it compiles on arxiv

\usepackage{graphicx} % Include figure files
\usepackage{dcolumn} % Align table columns on decimal point
\usepackage{xcolor}
\usepackage{bm} % bold math
\usepackage{physics}
\usepackage{amssymb,amsmath,amsfonts,mathtools}
\usepackage{amsthm}
\usepackage{bbm}
\usepackage{algorithm}
\usepackage[rightComments=false]{algpseudocodex}
\usepackage{hyperref}% add hypertext capabilities
\hypersetup{
    colorlinks,
    citecolor=blue,
    filecolor=blue,
    linkcolor=blue,
    urlcolor=blue
}
\usepackage[]{mdframed}
\mdfsetup{innerbottommargin=\topskip}

\newcommand{\pos}{\operatorname{pos}}

\begin{document}

\title{A Witness of GHZ Entanglement Using Only Collective Spin Measurements}

\author{Lin Htoo Zaw}
\affiliation{Centre for Quantum Technologies, National University of Singapore, 3 Science Drive 2, Singapore 117543}

\author{Khoi-Nguyen Huynh-Vu}
\affiliation{College of Engineering and Computer Science, VinUniversity, Gia Lam district, Hanoi 14000, Vietnam}

\author{Valerio Scarani}
\affiliation{Centre for Quantum Technologies, National University of Singapore, 3 Science Drive 2, Singapore 117543}
\affiliation{Department of Physics, National University of Singapore, 2 Science Drive 3, Singapore 117542} 

\begin{abstract}
Of existing entanglement witnesses that utilize only collective measurements of a spin ensemble, not all can detect genuine multipartite entanglement (GME), and none can detect Greenberger-Horne-Zeilinger (GHZ) states beyond the tripartite case. We fill this gap by introducing an entanglement witness that detects GME of spin ensembles, whose total spin is half-integer, using only collective spin measurements. Our witness is based on a nonclassicality test introduced by Tsirelson, and solely requires the measurement of total angular momentum along different directions. States detected by our witness are close to a family of GHZ-like states, which includes GHZ states of an odd number of spin-half particles. We also study the robustness of our witness under depolarizing noise, and derive exact noise bounds for detecting noisy GHZ states.
\end{abstract}

\maketitle

\section{Introduction}
The power of quantum technologies relies on the possibility of preparing superpositions of states of multipartite systems, i.e.,~entangled states \cite{guhne_toth_2009,metrology-review,review-nature}. As such, the detection of entanglement is crucial in demonstrating the ability of an experimental setup to produce such resource states with good fidelity. This paper contributes to the literature on the \textit{detection of multipartite entanglement}.

In multipartite systems, relying on local measurements or tomography becomes increasingly difficult, and quickly downright impossible, with the number of parties. Furthermore, in several experimental platforms, individual addressing of each particle might not be available. For these reasons, there have been many proposals of multipartite entanglement witnesses \textit{that require only the measurement of collective observables} like the center-of-mass position for a collection of material points, or the total magnetization for a collection of spins \cite{review-nature}. Here we consider spin ensembles, which appear naturally in several experimental architectures, like ultracold atoms in optical lattices \cite{optical-lattice-review}, spin defects \cite{spin-defects-review}, and donors in solid state materials \cite{spin-donors-review}. 

Entanglement witnesses for spin ensembles based on collective observables---usually, total angular momenta---are often built from generalized uncertainty relations, either taking the form of spin-squeezing inequalities or number-phase uncertainties \cite{spin_squeeze_extreme_2001,guhne_toth_2009, pezze_review_2018,sorensen_2001,korbicz_2005, korbicz_2006,number-phase-inequality}. They can detect certain families of multipartite-entangled states, like the Dicke and many-body singlet states \cite{toth_singlets_2004,toth_dicke_2007,toth_ssi_2007, toth_ssi_2009,vitagliano_sse_2011, vitagliano_sse_2014}.

But what type of entanglement can these witnesses detect? In multipartite systems, entanglement can take several forms: for instance, these quantum correlations can be present over some subdivisions of the system but not others. When the state is entangled over all possible bipartitions, it is said to have \textit{genuine multipartite entanglement (GME)} \cite{GME-coined}. As it turns out, among the witnesses cited above, only some can detect GME, and none can detect $N$-partite Greenberger-Horne-Zeilinger (GHZ) states beyond the tripartite case $N=3$. Since GHZ states are necessary in quantum protocols like distributed computing \cite{W-GHZ-resource} and secret-sharing \cite{GHZ-secret}, and are also useful for metrology \cite{GHZ-use-metrology} and error correction \cite{GHZ-use-error-correction}, it is desirable to find a witness that both utilizes only collective measurements and is able to detect GHZ states.

We fill this gap by introducing a GME witness that uses measurements of a collective observable (total angular momentum) to detect states GHZ states of any odd number of spin-half particles. The witness can be seen as ``dynamics-based''---a generalization of the precession protocol introduced by Tsirelson, which involves measuring the total angular momentum of the spin ensemble along a fixed direction at different times, under the assumption that the system is undergoing a uniform precession \cite{tsirelson_2006,Lin_2022}. But the assumption of dynamics is not necessary, and in this work we rather prescribe that the total angular momentum should be measured along different directions. GME is detected when the probability of obtaining a positive measurement outcome exceeds a separable bound that we compute explicitly.

\section{GHZ States, and the difficulty of witnessing them}

Consider a spin ensemble $\{j_n\}_{n=1}^N$ of $N$ particles, where the $n$th particle has spin $j_n$, and the total spin is $\sum_{n=1}^N j_n = K/2$ for odd $K$. The angular momentum vector of the $n$th particle is given by $\vec{J}^{(j_n)}$, while the total angular momentum of the spin ensemble is $\vec{J} = \sum_{n=1}^N\vec{J}^{(j_n)}$. We define a GHZ-like state to be
\begin{equation}
\ket{\text{GHZ}_K} \coloneqq \frac{1}{\sqrt{2}}\pqty{\bigotimes_{n=1}^N\ket{j_n,j_n} + \bigotimes_{n=1}^N\ket{j_n,-j_n}},
\end{equation}
which is exactly the GHZ state $\ket{\text{GHZ}_K} \propto \ket{\uparrow}^{\otimes K} + \ket{\downarrow}^{\otimes K}$ when all the particles are spin-half, so we shall loosely call this the GHZ state from here on out. Such a state has two important symmetries: it is an eigenstate of both $e^{-i\pi J_x/\hbar}$ and $e^{-i2\pi J_z /\hbar K}$.

The detection of GHZ entanglement is challenging because all its partial states with $L<K$ parties are the same as those of the incoherent mixture
\begin{equation}
    \rho=
    \frac{1}{2}\bigotimes_{n} \ketbra{j_n,j_n} +
    \frac{1}{2}\bigotimes_{n} \ketbra{j_n,-j_n}.
\end{equation} In other words, any witness of GHZ entanglement must involve $K$-body correlators. $K$ must therefore be known exactly---one cannot detect GHZ entanglement of an unknown number of spins. Among witnesses of multipartite entanglement, spin-squeezing inequalities and their extensions \cite{sorensen_2001,teh_reid_2019,toth_singlets_2004,li_gme_2021} rely on the violation of inequalities that involve variances of linear combinations of single-body observables. Other criteria, like the generalized spin-squeezing inequalities and energy-based witnesses \cite{toth_ssi_2007, toth_ssi_2009,vitagliano_sse_2011, vitagliano_sse_2014,toth_spin_2005}, rely on certain two-body correlators. As discussed, they necessarily fail to detect GHZ entanglement. Using uncertainty relations with three-body correlators, it was proved that $\ket{\text{GHZ}_3}$ can be witnessed using collective angular momentum measurements \cite{korbicz_2005,teh_reid_2019}; but those constructions were not extended beyond the tripartite case.

\section{The Protocol}
Our protocol is based on Tsirelson's precession protocol \cite{tsirelson_2006,Lin_2022}, but will be presented here without any reference to the dynamics (a later remark will connect this approach to the original dynamics-based protocol). Given a spin ensemble $\{j_n\}_{n=1}^N$ with $\sum_{n=1}^N j_n = K/2$ for odd $K$, each round of the protocol consists of:

(1) Initializing the spin ensemble.

(2) Drawing $k\in\{0,1,..., K-1\}$ with uniform probability $1/K$.

(3) Projectively measuring whether the total angular momentum in the direction 
\begin{eqnarray}
    J_k&\coloneqq& e^{-i (2\pi k/K) J_z/\hbar}\,J_x\,e^{i (2\pi k/K) J_z/\hbar}\nonumber\\
    &=&\cos(2\pi k/K)\,J_x\,+\,\sin(2\pi k/K)\,J_y
\end{eqnarray} is positive or negative.

After a large number of rounds, one estimates the probability $P_K$ of having found positive outcomes. For a generic state $\rho$ of the spin ensemble, this reads
\begin{equation}
\begin{aligned}
P_K &= \tr(\rho\, Q_K) & &\text{ with } & Q_K &\coloneqq \frac{1}{K}\sum_{k=0}^{K-1} \pos(J_k)\,,
\end{aligned}
\end{equation}
with the function $\pos$ being defined as $\pos(x > 0) = 1$, $\pos(x = 0) = 1/2$, and $\pos(x<0) = 0$.

It is clear that $Q_K$ shares the symmetries of the GHZ state: indeed, applying $e^{-i\pi J_x/\hbar}$ or $e^{-i 2\pi J_z /\hbar K}$ merely permutes the values of $k$ modulo $K$, which leaves $Q_K$ unchanged. To find the eigendecomposition of $Q_K$, we use the block decomposition $\vec{J} = \sum_{n=1}^N \vec{J}^{(j_n)} = \bigoplus_{j\in\mathcal{J}(\{j_n\}_{n=1}^N)} \vec{J}^{(j)}$, where $\mathcal{J}(\{j_n\}_{n=1}^N)$ is the set of irreducible spins that appear in the usual addition of angular momenta. As such, $Q_K$ permits the same block decomposition $Q_K \eqqcolon \bigoplus_{j\in\mathcal{J}(\{j_n\}_{n=1}^N)} Q_K^{(j)}$, where $Q_K^{(j)}$ is the observable of the protocol performed on a particle with a fixed spin $j$. The eigendecomposition of each $Q_K^{(j)}$ is known from Ref.~\cite{Lin_2022}:
\begin{equation}
    Q_K^{(j)} = \frac{1}{2}\mathbbm{1}^{(j)} + \begin{cases}
    0 & \text{if $j<K/2$}\\
    \begin{array}{l}
    \frac{1}{2^K}\binom{K-1}{\frac{K-1}{2}}\big(
    \ketbra{\mathbf{P}_{+K}}
    \\
    \quad{}-{}\ketbra{\mathbf{P}_{-K}} \big)
    \end{array}
    & \text{if $j=K/2$,}
    \end{cases}
\end{equation}
where $\ket{j,m}$ is the simultaneous eigenstate of $J_z$ and $\lvert\vec{J}\rvert^2$ with eigenvalues $\hbar m$ and $\hbar^2 j(j+1)$ respectively, while
\begin{align}
    \ket{\mathbf{P}_{\pm K}} &= \frac{1}{\sqrt{2}}\bqty{
        \ket{\tfrac{K}{2},\tfrac{K}{2}} \pm (-1)^{\frac{K-1}{2}}\ket{\tfrac{K}{2},-\tfrac{K}{2}}
    } \\
    &=\frac{1}{\sqrt{2}}\bqty{
        \bigotimes_{n=1}^N \ket{j_n,j_n} \pm (-1)^{\frac{K-1}{2}}\bigotimes_{n=1}^N \ket{j_n,-j_n}
    }\nonumber
\end{align} are GHZ-like states. Finally, summing over the blocks, we find
\begin{equation}\label{eq:QK-eigendecomposed}
    Q_K = \frac{1}{2}\bqty{\mathbbm{1} + \binom{K-1}{\frac{K-1}{2}} \frac{\ketbra{\mathbf{P}_{+K}}-\ketbra{\mathbf{P}_{-K}}}{2^{K-1}}}.
\end{equation}
Thus, the largest eigenvalue of $Q_K$,
\begin{equation}
    \mathbf{P}_K^{\textrm{max}}=\frac{1}{2}\bqty{1 + 2^{-(K-1)}\binom{K-1}{\frac{K-1}{2}}}\,,
\end{equation} is associated to the GHZ-like state $\ket{\mathbf{P}_{+ K}}$.

\section{Separable Bounds}
The original protocol was presented as a test of nonclassicality, as $P_K$ could not exceed $\mathbf{P}^c_K \coloneqq (1+1/K)/2$ if $\vec{J}$ were a classical vector \cite{Lin_2022}. This would be applicable to a single system. Since we are interested in entanglement, we are going to rather compute the bound 
\begin{equation}
   \mathbf{P}_K^{K\text{-sep}} \,\coloneqq\, \max_{\rho_{\text{sep}}}{\tr}(\rho_{\text{sep}}Q_K) 
\end{equation} over the set of separable states $\{\rho_{\text{sep}}\}$. A violation $P_K > \mathbf{P}_K^{K\text{-sep}}$ of this bound clearly implies that the system is entangled.

In general, the entanglement of a multipartite system depends on the chosen bipartition. For now, let us arbitrarily choose to partition $\{j_n\}_{n=1}^N$ into the subset $\mathbf{J} = \{j_{n_1},j_{n_2},\dots,j_{n_L}\}$ that contains $1 \leq L < N$ spins, and its complement $\mathbf{J}^\complement \coloneqq \{j_n\}_{n=1}^N \setminus \mathbf{J}$. Since $\sum_{l=1}^L \vec{J}^{(j_{n_l})} = \bigoplus_{j \in \mathcal{J}(\mathbf{J})}^{\sum_{l=1}^L j_{n_l}} \vec{J}^{(j)}$ by the addition of angular momenta, the Hilbert space of subsystem $\mathbf{J}$ is spanned by $\{\ket{j,m}\}_{j \in \mathcal{J}(\mathbf{J}),-j\leq m \leq j}$, where $\mathcal{J}(\mathbf{J})$ is the set of irreducible spins $j$ that appear in the block decomposition. Importantly, $j \leq \tilde\jmath \coloneqq \sum_{l=1}^L j_{n_l}$, where the subspace with spin $\tilde\jmath$ is nondegenerate. Similarly, the Hilbert space of $\mathbf{J}^\complement$ is spanned by $\{\ket{j',m'}\}_{j' \in \mathcal{J}(\mathbf{J}^\complement),-j'\leq m' \leq j'}$, and $j' \leq \tilde\jmath' \coloneqq K/2 - \tilde\jmath$ with nondegenerate $\tilde\jmath'$.

Under this choice, $\{\rho_{\text{sep}}\}$ are states separable over the bipartition $\mathbf{J}$-$\mathbf{J}^\complement$. Furthermore, since $\{\rho_{\text{sep}}\}$ is a convex set, we only need to maximize over its extremal points, which are the pure product states $\rho_{\text{sep}} = \lvert{\psi_\mathbf{J}}\rangle\!\langle{\psi_\mathbf{J}}\rvert \otimes \lvert{\psi_{\mathbf{J}^\complement}}\rangle\!\langle{\psi_{\mathbf{J}^\complement}}\rvert$, where $\lvert{\psi_\mathbf{J}}\rangle$ and $\lvert{\psi_{\mathbf{J}^\complement}}\rangle$ are states in $\mathbf{J}$ and $\mathbf{J}^\complement$ respectively. Hence, the separable bound is now given by
\begin{equation}\label{eq:sep-eigen}
\begin{aligned}
    \mathbf{P}_K^{K\text{-sep}} &=
    \max_{
        \lvert{\psi_\mathbf{J}}\rangle,
        \lvert{\psi_\mathbf{J}^\complement}\rangle
    } \pqty{
        \langle{\psi_\mathbf{J}}\rvert
        \otimes
        \langle{\psi_{\mathbf{J}^\complement}}\rvert
    } Q_K \pqty{
        \lvert{\psi_\mathbf{J}}\rangle
        \otimes
        \lvert{\psi_{\mathbf{J}^\complement}}\rangle
    } \\
    &= \max_{\lvert{\psi_\mathbf{J}^\complement}\rangle}
    \max_{\lvert{\psi_\mathbf{J}}\rangle} 
    \langle{\psi_\mathbf{J}}\rvert
    \tr_{\mathbf{J}^\complement}\!\bqty{
        Q_K\pqty{\mathbbm{1}^{(\mathbf{J})}
            \otimes
        \lvert{\psi_{\mathbf{J}^\complement}}\rangle
        \!\langle{\psi_{\mathbf{J}^\complement}}\rvert}
    }\lvert{\psi_\mathbf{J}}\rangle,
\end{aligned}
\end{equation}
where $\tr_{\mathbf{J}^\complement}$ is the partial trace over the $\mathbf{J}^\complement$ subsystem. As written, Eq.~\eqref{eq:sep-eigen} shows that $\mathbf{P}_K^{K\text{-sep}}$ can be found in a two-step process: first by evaluating the maximum eigenvalue of $\tr_{\mathbf{J}^\complement}\!\bqty{Q_K\pqty{\mathbbm{1}^{(\mathbf{J})} \otimes \lvert{\psi_{\mathbf{J}^\complement}}\rangle\!\langle{\psi_{\mathbf{J}^\complement}}\rvert}}$ for a fixed $\lvert{\psi_\mathbf{J}^\complement}\rangle$, then by maximizing this eigenvalue over all $\lvert{\psi_\mathbf{J}^\complement}\rangle$.

\begin{widetext}
So, let us parameterize $\lvert{\psi_\mathbf{J}^\complement}\rangle$ as $\lvert{\psi_\mathbf{J}^\complement}\rangle = \sum_{j' \in \mathcal{J}(\mathbf{J}^\complement)}\sum_{m'=-j'}^{j'} c_{j',m'} \ket{j',m'}$.
Substituting this into Eq.~\eqref{eq:QK-eigendecomposed}, we get
\begin{equation}
\tr_{\mathbf{J}^\complement}\!\bqty{
    Q_K\pqty{
        \mathbbm{1}^{(\mathbf{J})}
            \otimes
        \lvert{\psi_{\mathbf{J}^\complement}}\rangle
        \!\langle{\psi_{\mathbf{J}^\complement}}\rvert
    }
}
=
\frac{1}{2}\mathbbm{1}^{(\mathbf{J})} +
(-1)^{\frac{K-1}{2}}\frac{2^{-(K-1)}}{4}\binom{K-1}{\frac{K-1}{2}}
\sum_{\sigma \in \{\pm1\}} 
\sum_{\varsigma \in \{\pm1\}}
    c_{\tilde\jmath',\sigma\tilde\jmath'}^*
    c_{\tilde\jmath',\varsigma\tilde\jmath'}
    \ketbra{\tilde\jmath',\sigma \tilde\jmath'}{\tilde\jmath',\varsigma \tilde\jmath'}.
\end{equation}
Its maximal eigenvalue can be worked out to be
\begin{equation}
    \max_{\lvert{\psi_\mathbf{J}}\rangle} 
    \langle{\psi_\mathbf{J}}\rvert
    \tr_{\mathbf{J}^\complement}\!\bqty{Q_K\pqty{\mathbbm{1}^{(\mathbf{J})} \otimes \lvert{\psi_{\mathbf{J}^\complement}}\rangle
    \!\langle{\psi_{\mathbf{J}^\complement}}\rvert}}
    \lvert{\psi_\mathbf{J}}\rangle
    = \frac{1}{2}\bqty{ 1 + 2^{-K}\binom{K-1}{\frac{K-1}{2}}\pqty{\abs{c_{\tilde\jmath',\tilde\jmath'}}^2+\abs{c_{\tilde\jmath',-\tilde\jmath'}}^2}}.
\end{equation}
The upper bound $\abs{c_{\tilde\jmath',\tilde\jmath'}}^2 + \abs{c_{\tilde\jmath',-\tilde\jmath'}}^2 \leq \langle{\psi_\mathbf{J}}|{\psi_\mathbf{J}}\rangle = 1$ can be saturated with $c_{\tilde\jmath',m'} = \delta_{\tilde\jmath',m'}$, whence
\begin{equation}\label{eq:sep-bound}
    \mathbf{P}_K^{K\text{-sep}} =
    \frac{1}{2}\bqty{ 1 + 2^{-K}\binom{K-1}{\frac{K-1}{2}}}.
\end{equation}
Crucially, notice that Eq.~\eqref{eq:sep-bound} does not depend on the particular choice of $\mathbf{J}$-$\mathbf{J}^\complement$: in other words, $\mathbf{P}_K^{K\text{-sep}}$ is the separable bound for all bipartitions.\end{widetext}

Therefore, after performing the protocol on the spin ensemble, the observation $P_K > \mathbf{P}_K^{K\text{-sep}}$ certifies that it is entangled over all possible bipartitions---that is, the system is certified to be GME. Because of the gap
\begin{equation}\label{eq:dP}
   \mathbf{P}_K^{\textrm{max}}-\mathbf{P}_K^{K\text{-sep}}=2^{-(K+1)}\binom{K-1}{\frac{K-1}{2}},
\end{equation}
our witness can detect the GHZ-like state $\ket{\mathbf{P}_{+K}}$ as expected. In fact, any GHZ-like state $\propto \bigotimes_{n=1}^N\ket{j_n,j_n} + e^{i\phi}\bigotimes_{n=1}^N\ket{j_n,-j_n}$ with any phase $e^{i\phi}$ can be detected by the protocol with the replacement $J_k \to e^{-i\theta_i J_z/\hbar}J_ke^{i\theta_i J_z/\hbar}$, where
$\theta_i = [(K-1)\pi-2\phi]/(2K)$.

\section{Implementation of the Witness}
Let us provide a few comments regarding the implementation of the witness.

Firstly, the witness only requires measurements of $\pos(J_k)$, so the actual magnitude of the angular momentum does not need to be resolved. Hence, it is adequate to have an imprecise measurement apparatus that can only determine the sign of the total angular momentum. For example, measuring the total magnetization of a system gives information about the sign of its total angular momentum without needing to know the proportionality constant of the magnetic moment.
 
Secondly, if the spin system is evolving according to the Hamiltonian $H=-\omega J_z$, a dynamical implementation of the protocol becomes possible: one has to measure only $J_x$, the choice of direction becoming that of a time $t_k=(2\pi/\omega)k/K$ \cite{Lin_2022}. Even then, the requirement is that the dynamics be a uniform precession only over the duration of the certification process. As such, the dynamical implementation of our witness is particularly suited for certifying the entanglement of postquench states, which is a common problem in the study of spin ensembles \cite{igloi_toth_2023,quenchEW}. There, the spin ensemble would undergo some highly-entangling dynamics, after which the interactions are quickly turned off (``\textit{quenched}''), and the postquench system, now undergoing a uniform precession, would then be suitable for performing the precession protocol.

Thirdly, our witness can also be measured if the collective measurements occur not on the spin ensemble as a whole, but separately on different subensembles of the system, as was recently considered in Ref.~\cite{number-phase-inequality}. For example, take that each disjoint subensemble $\mathbf{J}_s \subseteq \{j_n\}_{n=1}^N$ can be collectively measured, where $\bigsqcup_s \mathbf{J}_s = \{j_n\}_{n=1}^N$. Then, $J_k = \sum_{s} \sum_{j_s \in \mathbf{J}_s} J_k^{(j_s)} \eqqcolon \sum_{s} J_k^{(\mathbf{J}_s)}$. As such, the protocol can be performed by collectively measuring $J_k^{(\mathbf{J}_s)} = \sum_{j_s \in \mathbf{J}_s} J_k^{(j_s)}$ of each subensemble $\mathbf{J}_s$, then classically postprocessing the measurement outcomes to obtain $\pos(\sum_s J_k^{(\mathbf{J}_s)} )$ at each time. Of course, all the spins must eventually be measured.

Lastly, while our witness can, in principle, detect GHZ-like states in any spin ensemble with a total spin that is a finite half-integer, it fails in the thermodynamical limit. This comes from the fact that the gap \eqref{eq:dP} scales as $\sim K^{-\frac{1}{2}}$, which disappears when $K\to\infty$. This scaling also means that the applicability of our witness for large spin ensembles would, in practice, depend on the accuracy of the measurement apparatus.

\section{Robustness Under Depolarizing Noise}
Finally, we study the robustness of the protocol for detecting GHZ-like states in the presence of depolarizing noise. Depolarizing noise is given by the linear map
\begin{equation}
    \Lambda(\bullet) = p \frac{\mathbbm{1}}{\tr(\mathbbm{1})} + (1-p)\bullet
\end{equation}
which describes the replacement, with probability $p$, of the input state with the maximally-mixed state.

We will consider both global and local depolarizing noise. The global depolarizing channel $\Lambda_G$ acts on the spin ensemble as a whole, while the local depolarizing channel is given by $\Lambda_L \coloneqq \bigotimes_{n=1}^N \Lambda_{n}$, where each $\Lambda_n$ acts individually on the $n$th particle.

For the global case, let $P_K^{(G)}$ be the score obtained by $\ket{\mathbf{P}_{+K}}$ in the presence of global depolarizing noise with parameter $p_G$. Then,
\begin{equation}
\begin{aligned}
    P_K^{(G)} &=\tr[\Lambda_G(\ketbra{\mathbf{P}_{+K}})Q_K] \\
    &= p_G\frac{\tr(Q_K)}{\tr(\mathbbm{1})} +
    (1-p_G)\bra{\mathbf{P}_{+K}}Q_K\ket{\mathbf{P}_{+K}} \\
    &= \frac{1}{2} + 2(1-p_G) \pqty{\mathbf{P}_{K}^{K\text{-sep}}-\frac{1}{2}}.
\end{aligned}
\end{equation}
Our witness certifies GME of noisy GHZ-like states when $P_K^{(G)} > \mathbf{P}_{K}^{K\text{-sep}}$, and hence is detected for the range $p_G < 1/2$. It is known that GHZ states are GME in the presence of global depolarizing noise if and only if $p_G < 1/[2(1-2^{-K})]$ \cite{mixed-GHZ-max-GME}, so our protocol detects noisy GHZ states close to the theoretical limit.
\begin{widetext}
Meanwhile, for local depolarizing noise with noise parameters $\{p_n\}_{n=1}^N$, we first work out a preliminary step
\begin{equation}\label{eq:local-noise-step}
\begin{aligned}
    \bra{\mathbf{P}_{\pm K}}\Lambda_{L}\pqty\big{\ketbra{\mathbf{P}_{+K}}}\ket{\mathbf{P}_{\pm K}} &= \frac{1}{2}\bra{\mathbf{P}_{\pm K}}\Bigg(
        \bigotimes_{n=1}^N\bqty\Big{ \Lambda_n\pqty\big{\ketbra{j_n,-j_n}} } +
        \bigotimes_{n=1}^N\bqty\Big{ \Lambda_n\pqty\big{\ketbra{j_n,j_n}} } \\
    &\qquad{}+{}
        (-1)^{\frac{K-1}{2}}\Bqty{\bigotimes_{n=1}^N\bqty\Big{ \Lambda_n\pqty\big{\ketbra{j_n,-j_n}{j_n,j_n}} } +
        \bigotimes_{n=1}^N\bqty\Big{ \Lambda_n\pqty\big{\ketbra{j_n,-j_n}{j_n,-j_n}} }
        }
    \Bigg)\ket{\mathbf{P}_{\pm K}} \\
    &= \frac{1}{2}\bqty{
        \prod_{n=1}^N\pqty{\frac{p_n}{\tr(\mathbbm{1}^{(j_n)})}+1-p_n}
        \prod_{n=1}^N\pqty{\frac{p_n}{\tr(\mathbbm{1}^{(j_n)})}}
        \pm \prod_{n=1}^N\pqty{1-p_n}
    }.
\end{aligned}
\end{equation}
\end{widetext}
With Eq.~\eqref{eq:local-noise-step}, the score $P_K^{(L)}$ obtained by $\ket{\mathbf{P}_{+K}}$ under the presence of local depolarizing noise can be worked out to be
\begin{equation}
\begin{aligned}
    P_K^{(L)} &= \tr[\Lambda_{L}\pqty\big{\ketbra{\mathbf{P}_{+K}}} Q_K] \\
    &= \frac{1}{2} + 2 \prod_{n=1}^N\pqty{1-p_n} \pqty{\mathbf{P}_{K}^{K\text{-sep}}-\frac{1}{2}}.
\end{aligned}
\end{equation}
As such, our witness detects GME of noisy GHZ-like states in the presence of local depolarizing noise when the noise parameters are in the range $\prod_{n=1}^N\pqty{1-p_n} > 1/2$. If the local noise is identical for all particles such that $p_n = p_L$ for all $n$, this condition becomes $p_L < 1-2^{-\frac{1}{N}}$.

Therefore, our witness is robust against depolarizing noise when it acts globally on the system, but is less robust when it acts locally on each particle.

\section{Conclusion}
We introduced a witness of GHZ-like entanglement that uses only collective measurements of the total spin angular momentum for spin ensembles whose total spin is a half-integer. This result fills a gap in the body of work of entanglement witnesses that utilize collective measurements, as existing techniques based on uncertainty relations or total energy measurements cannot detect GHZ-like states beyond the tripartite case.

Theoretically, our witness can detect GHZ-like entanglement for any spin ensemble whose total spin is a finite half-integer. Practically, how large of a spin ensemble the witness can detect is dependent on experimental factors. Precision certainly matters, since the gap between the separable bound and the score of the GHZ-like state decreases as the size of the spin ensemble increases. Another factor is the type of noise present: as we showed in the robustness study, our witness is more robust to depolarizing noise if it acts globally rather than locally on the system.

Our witness can also be extended in several ways. When performing the protocol, one could replace $\pos(J_x) \to f_0\mathbbm{1} + f_{\text{odd}}(J_x)$ for an odd function $f_{\text{odd}}(-x) = - f_{\text{odd}}(x)$. Then, a similar analysis gives the separable bound $f_0 + f_K/2$ with $f_K \coloneqq \lvert \langle \frac{K}{2},\frac{K}{2} \rvert f_{\text{odd}}(J_x) \lvert \frac{K}{2}, -\frac{K}{2} \rangle \rvert$ for $K$ odd. This means that GHZ states will be detected by this modified witness as long as $f_K > 0$, which requires $\frac{d^{K+2L}}{dx^{K+2L}}f_{\text{odd}}(x) \rvert_{x=0} \neq 0$ for some integer $L \geq 0$. Of course, by choosing $f_{\text{odd}}(x) \propto 2\pos(x)-1$  as we have done here, the magnitudes of the angular momentum measurements do not need to be resolved, which is a benefit that might be lost with other choices of $f_{\text{odd}}(x)$.

Another possible extension is to perform the witness for a spin ensemble whose total spin $\sum_{n=1}^N j_n \neq K/2$ does not match the number of measurements, which we consider in the longer companion paper \cite{This-Paper}. There, we find that the separable bound increases, and that the witness is able to detect other GME states, which are different from GHZ states but share similar symmetry properties, that are also missed by previously proposed witnesses.

We conclude with some open problems. The most evident is the failure of our witness to detect GHZ-like states when the total spin of the ensemble is integer-valued. Due to the symmetry of the protocol, a trivial score will be obtained for every state when an even number of measurements are made. This symmetry has to be broken to detect GHZ-like entanglement when the total spin is an integer, or perhaps a very different protocol has to be used altogether.

Another open problem is the scaling of the gap between the score of the GHZ-like state and the separable bound. The current scaling is such that GHZ-like states up to $K \lesssim 20$ can be detected if a $\sim 5\%$ difference in probability can be resolved, and up to $K \lesssim 400$ if a $\sim 1\%$ difference can be resolved. An improved scaling can increase the size of the spin ensemble in which the witness can be used in practice, and would be more robust against noise and imperfections.

\section*{Acknowledgments}
We would like to thank G\'eza T\'oth for his detailed comments on an early draft of this manuscript. Hung Nguyen Quoc is thanked by K.N.H.V.~for his mentorship, and by V.S.~for his hospitality at VNU, Hanoi.

This work is supported by the National Research Foundation, Singapore, and A*STAR under its CQT Bridging Grant. 

\bibliography{references}

\end{document}